\documentclass[12pt,preprint,nofootinbib]{revtex4-1}
\usepackage{amssymb,amsmath,amscd}
\usepackage{amsfonts,latexsym,bm}
\usepackage[colorlinks=true,linkcolor=blue,citecolor=blue,urlcolor=blue]{hyperref}
\newcommand{\be}{\begin{equation}}
\newcommand{\ee}{\end{equation}}
\newcommand{\bea}{\begin{eqnarray}}
\newcommand{\eea}{\end{eqnarray}}
\newcommand{\p}{\partial}
\newcommand{\nn}{\nonumber}
\newcommand{\cA}{\mathcal{A}}

\begin{document}



\title{Consistent mass formulas for higher even-dimensional Reissner-Nordstr\"{o}m-NUT-AdS
spacetimes}

\author{Shuang-Qing Wu}
\email{sqwu@cwnu.edu.cn}

\author{Di Wu}
\email{wdcwnu@163.com}
\affiliation{School of Physics and Astronomy, China West Normal University,
Nanchong, Sichuan 637002, People's Republic of China}

\date{\today}

\begin{abstract}
In our previous work [D. Wu and S.-Q. Wu, \href{https://doi.org/10.1103/PhysRevD.108.064034}{Phys.
Rev. D \textbf{108}, 064034 (2023)}], we have formulated consistent thermodynamic first law and 
Bekenstein-Smarr mass formulas for all higher even-dimensional uncharged Taub-NUT (AdS) spacetimes 
through the inclusion of a secondary hair $J_n = Mn$ as did in the four-dimensional NUT-charged 
cases [S.-Q. Wu and D. Wu, \href{https://doi.org/10.1103/PhysRevD.100.101501}{Phys. Rev. D 
\textbf{100}, 101501(R) (2019)}]. In this article, we will show that the corresponding 
Reissner-Nordstr\"{o}m(RN)-NUT (AdS) cases in all higher even-dimensions can be coherently treated 
just like the uncharged cases if another secondary hair $Q_n \propto{} qn$ is further introduced as 
did in the four-dimensional dyonic NUT-charged case [D. Wu and S.-Q. Wu, 
\href{https://doi.org/10.1103/PhysRevD.105.124013}{Phys. Rev. D \textbf{105}, 124013 (2022)}]. 
As a by-product of our consequence, it is also demonstrated that there is no need to introduce the 
secondary hair $Q_n$ in the mass formulas for the four-dimensional purely electric-charged RN-NUT 
(AdS) spacetimes.
\end{abstract}

\maketitle

\section{Introduction}

Recently, there has been a considerable interest \cite{PRD100-101501,PRD105-124013,2210.17504,
PRD100-064055,JHEP0719119,CQG36-194001,PLB798-134972,PRD100-104016,JHEP0520084,JHEP1022044,
IJMPD31-2250021,PLB802-135270,PRD106-024022,JHEP0321039,PRD101-124011,PRD105-124034,JHEP1022174}
in formulating the consistent thermodynamics of the four-dimensional Lorentzian Taub-NUT
spacetimes. According to our viewpoint \cite{2209.01757}, these attempts can be roughly
classified into three different schemes: (I) The ``multi-hair" formalism \cite{PRD100-101501,
PRD105-124013}, which is a most perfect generalization of the traditional black hole thermodynamics.
In this formulation, an extra secondary hair $J_n = mn$ (and ($qn, pn$) for the dyonic case)
is introduced, the first law and Bekenstein-Smarr formula are derived from the squared-mass,
which can be thought of as representing a hyper-surface in one higher dimensional thermodynamic
state space, the well-known Bekenstein-Hawking one-quarter area-entropy law $S = A/4$ and
Hawking-Gibbons temperature $T = \kappa/(2\pi)$ are the natural consequences of our algebraic
manipulation. (II) The ``$\psi-\mathcal{N}$" formalism, where a new nonglobal Misner charges
is imported \cite{PRD100-064055,JHEP0719119,CQG36-194001,PLB798-134972,PRD100-104016,JHEP0520084,
JHEP1022044}; and the ``$J_n-\Xi$" formalism \cite{IJMPD31-2250021} which is related to the
former one via the decomposition: $\Xi = 2J_n-\mathcal{N}/(4\pi)$. (III) The ``modified mass"
formalism \cite{PRD101-124011,PRD105-124034}, where the thermodynamic mass appeared in the mass
formulas is the internal energy that receives from the contribution of the Misner strings. Via a
Legendre transformation, this formulation transfers to the ``horizon mass'' formalism  advocated
in \cite{JHEP1022174}. In the formalisms (II) and (III), the relations $S = A/4$ and $T =
\kappa/(2\pi)$ are assumed to hold true from the beginning.

Given the great success of these researches in the four-dimensional NUT-charged spacetimes, one
can anticipate that the higher even-dimensional NUT-charged spacetimes should also own similar
properties, namely, they obey similar first law and Bekenstein-Smarr mass formula. Then, a naive
question is whether these existing methods can be extended to deal with the thermodynamics of
the higher even-dimensional NUT-charged spacetimes. However, the answer seems to be not so
optimistic. At present, there is only one work \cite{2209.01757} that conducted such a task,
and all the remaining efforts are only restricted to the four-dimensional NUT-charged cases.
It is now clear that any technique \cite{PRD100-064055,JHEP0719119,CQG36-194001,PLB798-134972,
PRD100-104016,JHEP0520084,JHEP1022044,JHEP1022174} that was based upon the Komar integral is
only successful in the treatment of the four-dimensional NUT-charged spacetimes, and will be
foredoomed to fail in the higher even-dimensional NUT-charged cases, simply because the Komar
integral that is used to calculate the mass of these ($D > 4$) even-dimensional NUT-charged
spacetimes is divergent at spatial infinity. The same is also true for the completion method
that was adopted to compute the mass of the higher even-dimensional NUT-AdS spacetimes
\cite{2209.01757}. Both approaches are merely lucky in the four-dimensional NUT-charged
spacetime, and become invalid in the ($D > 4$) even-dimensional NUT-charged cases. To regulate
the divergency at spatial infinity so as to get a finite expression, one must subtract this
divergency due to the contribution from the massless purely NUT background \cite{2209.01757}.
On the other hand, the Abbott-Deser (AD) method \cite{NPB195-76,PRD73-064020} can be used to
get the finite expressions for the conserved mass of the arbitrary even-dimensional NUT-charged
spacetimes both with and without a nonzero cosmological constant.

Here we would like to especially point out that the formal derivation of the Bekenstein-Smarr
mass formula delivered in Ref. \cite{CQG36-194001} for all dimensional NUT-AdS spacetimes, which
is based upon the Komar integral with the removal of the divergence by using the Killing-Yano
tensor, only works effectively in the four-dimensional NUT-AdS cases, but is invalid in the
($D > 4$) even-dimensional NUT-AdS case. Therefore, the higher even-dimensional NUT-charged
solutions provide a perfect touchstone to test the universality of all the above-mentioned methods.
In our opinion, apart from our ``multi-hair" prescription, we anticipate that the ``modified
mass" formalism \cite{PRD101-124011,PRD105-124034} which utilizes the thermodynamic Gibbs-Duhem
relation, might be another feasible method to study thermodynamics of the Lorentzian higher
even-dimension Taub-NUT spacetimes. Nevertheless, our motivation of the present article is
to generalize our previous work \cite{2209.01757} to the higher even-dimension RN-NUT (AdS)
spacetimes \cite{CQG23-2849,PLB632-537,PRD73-124039}, and to show that our ``multi-hair" method
is also a very effective tool.

On the other hand, by virtue of AdS/CFT correspondence, the Euclidean sectors of the higher
even-dimensional Taub-NUT/bolt spacetimes \cite{hepth9809041,CQG19-2051,PLB593-218} had been
extensively investigated during the first decade of this century in different contexts \cite{
PRD60-104001,PRD72-083032,PLB620-1,NPB652-348,NPB674-329,PRL91-061031,IJMPA19-3987,JHEP0105049,
JHEP0306083,PLB738-294}, especially in the thermodynamical aspects \cite{PRD72-083032,NPB652-348,
JHEP0306083,PLB738-294}. The most remarkable aspect of these researches is to reveal that the
NUT-charged AdS and dS spacetimes own a lot of rather peculiar, even unusual properties. Of
particular interest is to note that the counterterm method \cite{PRD60-104001,NPB652-348} must
be adopted to obtain the finite expressions of the Euclidean action and the thermodynamic mass,
whilst the conformal completion method is not enough to get a finite result, some additional
background subtractions still need to be supplemented \cite{JHEP0306083}.

In this paper, we purpose to extend our previous work \cite{2209.01757} to the charged cases
of all higher even-dimension RN-NUT (AdS) spacetimes \cite{CQG23-2849,PLB632-537,PRD73-124039}.
We first present the explicit solutions and determine all the thermodynamic quantities that
can be computed by the standard technique, especially the conserved mass by the AD procedure
\cite{NPB195-76}. In particular, we will show that neither the Komar integral nor its modified
form at spatial infinity can give a finite expression for the mass of all $D > 4$ dimensional
NUT-charged (AdS) spacetimes. Then in the remaining two sections, we will separately discuss
the consistent thermodynamics of the higher even-dimension RN-NUT spacetimes without and with
a negative cosmological constant. To do so, we first deduce the squared-mass formulas for both
cases by not only introducing a secondary hair $J_n = Mn$ just like the uncharged cases
\cite{2209.01757}, but also adding another secondary hair $Q_n \propto{} qn$ as did in the
four-dimensional dyonic NUT-charged case \cite{PRD105-124013}. These squared-mass formulas
are then regarded as representing the hyper-surfaces embedded into two more high dimensional
thermodynamic state spaces, from which one can derive the first law via the simple differentiation
with respect to their state variables and obtain their corresponding conjugate potentials.
Subsequently, it can be checked that all these conjugate pairs not only perfectly satisfy
the extended Bekenstein-Smarr mass formula, but also obey the common Maxwell relations. By
the way, it will also be demonstrated that there is no need to introduce the secondary hair
$Q_n = Qn = qn$ in the mass formulas for the four-dimensional purely electric-charged RN-NUT
(AdS) spacetimes. Finally, our article is ended up by presenting our conclusions and outlooks.
In the Appendix \ref{appe}, we devise another universal method to compute the primary (the mass)
and secondary hairs: $J_n$ and $Q_n$.

\section{The ($2k+2$)-dimensional RN-NUT (AdS) spacetimes}

The Lagrangian of arbitrary dimensional Einstein-Maxwell gravity theory with a negative
cosmological constant is
\be
\mathcal{L} = \sqrt{-g}\big[R +2k(2k+1)g_0^2 -F^2\big] \, ,
\ee
where $F^2 = F_{ab}F^{ab}$, and the field strength 2-form is $F = d\mathbb{A}$ with respect
to the Abelian 1-form potential $\mathbb{A}$. The field equations derived from the action
read
\be
R_{ab} +(2k+1)g_0^2g_{ab} = 2F_{ac}F_b^{~c} -\frac{1}{2k}g_{ab}F^2 \, , \qquad
\nabla_b{}F^{ab} = 0 \, .
\ee

The exact solutions of the ($2k+2$)-dimensional Lorentzian RN-NUT AdS spacetimes for a
U(1) fibration over the base manifold $\otimes_{i=1}^k{}S^2$ are compactly written as
\cite{PLB632-537}

\bea
&& ds_{2k+2}^2 = -f(r)\Big(dt +2n\sum_{i=1}^k\cos\theta_i{}d\phi_i\Big)^2
 +\frac{dr^2}{f(r)} \nn \\
&&\qquad\qquad +\big(r^2 +n^2\big)\sum_{i=1}^k\big(d\theta_i^2
 +\sin^2\theta_i{}d\phi_i^2\big) \, , \label{hdNUT} \\
&&\qquad \mathbb{A} = \frac{\sqrt{k}\, qr}{\big(r^2 +n^2\big)^k}\Big(dt
 +2n\sum_{i=1}^k\cos\theta_i{}d\phi_i\Big) \, , \label{mathbbA}
\eea
with the radial structure function being
\bea
&& f(r)= \frac{r}{\big(r^2+n^2\big)^k}\int^r\big[1 +(2k+1)g_0^2
 \big(x^2+n^2\big)\big]\frac{\big(x^2+n^2\big)^k}{x^2}dx \nn \\
&&\qquad\quad -\frac{2mr}{\big(r^2+n^2\big)^k}
 +q^2\frac{(2k-1)r^2 +n^2}{\big(r^2+n^2\big)^{2k}} \, .
\eea
in which, three parameters ($m, n, q$) are related to the mass, the NUT charge and the electric
charge of the RN-NUT AdS solutions, respectively, while $g_0 = 1/l$ is the gauge coupling constant
that is identical to the reciprocal of the cosmological scale. In the above, we had also made
such a gauge choice that the temporal component of the Abelian potential $\mathbb{A}$ vanishes
at infinity.

We begin with by determining some thermodynamic quantities that can be evaluated via the
standard method. Firstly, the horizon area $A_h$ and the surface gravity $\kappa = f^{\prime}
(r_h)/2$ at the Killing horizons specified by $f(r) = 0$ are easily computed as
\bea
&&A_h = \big[4\pi\big(r_h^2+n^2\big)\big]^k = \big(4\pi\big)^k\cA \, , \\
&&\kappa = \frac{1 +(2k+1)g_0^2\big(r_h^2+n^2\big)}{2r_h}
 -q^2\frac{\big(r_h^2+n^2\big)^2 +4k(k-1)r_h^4}{2r_h\big(r_h^2+n^2\big)^{2k+1}} \, ,
\label{kath}
\eea
where a reduced area: $\cA = \big(r_h^2+n^2\big)^k$ is introduced for latter convenience as
did in \cite{2209.01757}, in which $r_h$ represents the greatest root of the horizon equation:
$f(r_h) = 0$.

Secondly, the electric charge $Q$ measured at spatial infinity can be computed by using the
Gauss' law integral
\be
Q = -\frac{(2\pi)^k}{4\pi}\prod_{i=1}^k\int_0^{\pi}d\theta_i
 \big(\sqrt{-g}F^{tr}\big)\big|_{r\to\infty}
 =\sqrt{k}(2k-1)(4\pi)^{k-1}q \, ,
\ee
while the same integral at the horizon gives
\be
Q_h = \sqrt{k}(4\pi)^{k-1}q \frac{(2k-1)r_h^2-n^2}{r_h^2+n^2} \, .
\ee
The corresponding electrostatic potential is gauge independent by virtue of the above gauge
choice, and is simply given by
\be
\Phi = \mathbb{A}_t(r_h) = \frac{\sqrt{k}\, qr_h}{\big(r_h^2 +n^2\big)^k} \, .
\ee

Thirdly, we will now show that neither the Komar integral nor its modified form at spatial
infinity can give a finite expression for the mass in all $D > 4$ dimensional NUT-charged
spacetimes. The usual Komar integral associated with the timelike vector $\xi^a = (\p_t)^a$ is:
\bea
&& \mathcal{M}(r) = -\frac{(2\pi)^k}{4\pi}\prod_{i=1}^k\int_0^{\pi}d\theta_i
 \big(\sqrt{-g}\xi^{[t;r]}\big)\big|_{r\to\infty}
= \frac{1}{2}(4\pi)^{k-1}\big(r^2+n^2\big)^k\p_r{}f(r) \nn \\
&&\qquad~ = (2k-1)(4\pi)^{k-1}m  +(4\pi)^{k-1}g_0^2r\big(r^2+n^2\big)^k
 -2(4\pi)^{k-1}\big[1 \nn \\
&&\qquad\qquad +2(k+1)g_0^2n^2\big]\sum_{j=0}^{k-1}\sum_{i=0}^{j-1}
 \frac{(-1)^i\,k!n^{2k-2j+2i}r^{2j-2i-1}}{(4j^2-1)\,j!(k-j-1)!}
  +\mathcal{O}(r^{-1}) \, , \qquad \label{Mr}
\eea
which is clearly divergent at spatial infinity in all $D > 4$ dimensions.

In Ref. \cite{CQG36-194001}, the modified Komar integral is defined via the replacement
of $\xi^{[a ; b]}$ by
\be
\tilde{\xi}^{ab}\equiv \xi^{a ; b} -\xi^{b ; a} +2g_0^2\mathbf{k}^{ab}
= \frac{-1}{2k}C^{abcd}\mathbf{k}_{cd} \, , \label{KY}
\ee
where $C^{abcd}$ is the Weyl tensor, $\mathbf{k} = d\mathbf{b}$ is the second-order
Killing-Yano tensor generated by
\be
\mathbf{b} = \frac{1}{2}(r^2+n^2)\Big(dt +2n\sum_{i=1}^k\cos\theta_i{}d\phi_i\Big) \, ,
\ee
and satisfies $\mathbf{k}^{ab}_{~~;b} = (2k+1)\xi^a$. In the last identity of Eq. (\ref{KY}),
we also show that the modified Komar potential is related to the Y-ADM potential proposed in
Ref. \cite{JHEP0804045}.

Replaced by the modified Komar potential $\tilde{\xi}^{ab}$, the mass integral becomes
\bea
&&\widetilde{M}(r) = (2k-1)(4\pi)^{k-1}m -2(4\pi)^{k-1}\big[1 +2(k+1)g_0^2n^2\big] \nn \\
&&\qquad\qquad \times\sum_{j=0}^{k-1}\sum_{i=0}^{j-1}
 \frac{(-1)^i\,k!n^{2k-2j+2i}r^{2j-2i-1}}{(4j^2-1)\,j!(k-j-1)!} +\mathcal{O}(r^{-1}) \, ,
\eea
in which the second term still can not be regulated away and explicitly reads
\bea
&& 0 \, , \quad \frac{-16\pi}{3}n^2r\big(1 +6g_0^2n^2\big) \, , \quad
 \frac{-32\pi^2}{5}\big(1 +8g_0^2n^2\big)n^2r\big(r^2 +9n^2\big) \, , \nn \\
&&\quad \frac{-512\pi^3}{35}\big(1 +10g_0^2n^2\big)n^2r\big(r^4 +6n^2r^2 +29n^4\big) \, , \nn
\eea
in the $D = 4, 6, 8, 10$ dimensions, respectively.

Consequently, it has clearly shown that the formal derivation of the Bekenstein-Smarr mass
formula conducted in Ref. \cite{CQG36-194001} for all even-dimensional NUT-AdS spacetimes
is invalid in all ($D > 4$) even-dimensional NUT-AdS cases, but merely works effectively in
the four-dimensional NUT-AdS cases. It can also be inferred that all prescriptions that are
based upon the (modified) Komar integral can not be directly applied to all the cases of the
($D > 4$) even-dimensional NUT-charged (AdS) spacetimes. The same situation also occurs
\cite{2209.01757} when one adopted \cite{JHEP0306083} the conformal completion method, which
is successful only fortunately in the four-dimensional NUT-AdS cases. Both methods must be
complemented with some additional reference background subtractions for the divergent terms.

Alternatively, there are two different approaches available in the literatures to get a finite
expression for the mass of the NUT-charged spacetimes in all even dimensions: The counterterm
method \cite{PRD60-104001,NPB652-348} and the AD method \cite{NPB195-76,PRD73-064020}. The
former is universal in the sense of its background independent, but the related computation
is very cumbersome, while the latter is a reference background subtraction approach, and due
to its dimensional independent, the application of the AD procedure is quite easy and suitable
for all dimensions. Below we will employ this approach to calculate the mass of the higher
even-dimensional NUT-charged (AdS) spacetimes. In this procedure, the AD mass is a global
conserved charge defined for spacetimes with an arbitrary asymptotic behavior and is associated
with the isometry of the asymptotic geometry.

For our proposal, the spacetime metric is decomposed into a perturbative form: $g_{ab} =
\bar{g}_{ab} +h_{ab}$, where the massless background line element  $d\bar{s}^2 = \bar{g}_{ab}
dx^{a}dx^{b}$ is chosen to be the pure NUT-charged metric. The background metric is still
represented by Eq. (\ref{hdNUT}) only with the radial function being replaced by $f_0(r)$,
which is obtained by setting $m = q = 0$ in $f(r)$:
\bea
f_0(r) = f(r)\big|_{m=q=0} = \frac{r}{\big(r^2+n^2\big)^k}\int^r\big[1 +(2k+1)g_0^2
 \big(x^2+n^2\big)\big]\frac{\big(x^2+n^2\big)^k}{x^2}dx  \, , \nn
\eea
or simply by setting $m=0$ in $f(r)$. Accordingly, the background determinant is: $\sqrt{-
\bar{g}} = \sqrt{-g} = (r^2+n^2)^k\prod_{i=1}^k\sin\theta_i$. Thus, the perturbation part
is precisely given by:
\bea
h_{ab}dx^{a}dx^{b} = \big(f_0(r) -f(r)\big)\Big(dt +2n\sum_{i=1}^k
 \cos\theta_i{}d\phi_i\Big)^2 +\frac{f_0(r) -f(r)}{f_0(r)f(r)}dr^2 \, , \nn
\eea
via the background substraction, and one can easily get every components of the metric tensor
$h_{ab}$, which asymptotically vanishes at spatial infinity.

To apply the AD procedure, one first needs to define a super-potential:
\be
\mathbb{K}^{abcd} =\bar{g}^{ac}\tilde{h}^{bd} +\bar{g}^{bd}\tilde{h}^{ac}
 -\bar{g}^{ad}\tilde{h}^{bc} -\bar{g}^{bc}\tilde{h}^{ad} \, , \qquad
\tilde{h}^{ab} = h^{ab} -(1/2)\bar{g}^{ab}h^c_{~c} \, ,
\ee
with all indices being raised and lowered by using the background metric tensors $\bar{g}^{ab}$
and $\bar{g}_{ab}$. Furthermore, one can define a symmetric tensor \cite{NPB195-76}:
\be
\mathbb{Q}^{ab} = \xi_c\bar{\nabla}_d{}\mathbb{K}^{abcd}
 +\mathbb{K}^{acdb}\bar{\nabla}_d\xi_c \, ,
\ee
so that the AD mass, which acts as the conserved charge associated with the timelike Killing
vector $\xi^a = \delta_t^a$, can be calculated via the following integral at infinity:
\be \label{ADm}
\mathcal{Q}[\xi] = \frac{(2\pi)^k}{16\pi}\prod_{i=1}^k\int_0^{\pi}d\theta_i\,
 \big(\sqrt{-\bar{g}}\, \mathbb{Q}^{tr}\big)\big|_{r\to\infty} \, .
\ee

For the ($2k+2$)-dimensional NUT-charged AdS solutions, the integrand in the above expression
behaves like:
\be
\sqrt{-\bar{g}}\, \mathbb{Q}^{tr} \sim 4k\, m\prod_i^k\sin\theta_i +\mathcal{O}(r^{-2}) \, ,
\ee
for all $(D > 4)$ dimensional cases ($k > 1$); when $k=1$, its asymptotic expansion at spatial
infinity is: $(4m -2q^2/r)\sin\theta +\mathcal{O}(r^{-2})$. Thus, one can straightforwardly get
the conserved AD mass for all even-dimensional NUT-charged AdS spacetimes as follows:
\be
M = \mathcal{Q}[\p_t] = k(4\pi)^{k-1}m \, .
\ee
Via the analogy with the electric mass $M$, we will simply take the expression of the NUT charge
as
\be
N = k(4\pi)^{k-1}n \, .
\ee

\section{Mass formulas of ($2k+2$)-dimensional RN-NUT spacetimes}

In this section, we will first study a simpler case in which the gauge coupling
constant $g_0$ vanishes. The metric and the Abelian gauge potential are given by
Eqs. (\ref{hdNUT}-\ref{mathbbA}), but now the radial function is
\be
f(r) = \frac{r}{\big(r^2+n^2\big)^k}\bigg\{\int^r \frac{\big(x^2+n^2\big)^k}{x^2}dx
 -2m\bigg\} +q^2\frac{(2k-1)r^2 +n^2}{\big(r^2+n^2\big)^{2k}} \, . \label{fg0}
\ee

Let us simply collect all the thermodynamic quantities that have been computed in the
last section. The horizon area, the surface gravity and the electrostatic potential at
the horizon are
\be
A_h = \big(4\pi\big)^k\cA \, , \quad
\kappa = \frac{1}{2r_h} -q^2\frac{\big(r_h^2+n^2\big)^2
 +4k(k-1)r_h^4}{2r_h\big(r_h^2+n^2\big)^{2k+1}} \, , \quad
\Phi = \frac{\sqrt{k}\, qr_h}{\big(r_h^2 +n^2\big)^k} \, .
\label{kappag0}
\ee
The AD mass, the NUT charge $N$ and the electric charge are
\be
M = k(4\pi)^{k-1}m \, , \quad  N = k(4\pi)^{k-1}n \, , \quad
Q =\sqrt{k}(2k-1)(4\pi)^{k-1}q \, .
\ee
These conserved charges ($M$, $N$, $Q$) are the primary hairs that explicitly appear
in the leading order of the asymptotic expansions of the following components of the
metric and the Abelian potentials at infinity:
\bea
&& g_{tt} \simeq -\frac{1}{2k-1} +\frac{2m}{r^{2k-1}} +\cdots \, , \qquad
\mathbb{A}_t \simeq \sqrt{k}\frac{q}{r^{2k-1}} +\cdots \, , \nn \\
&&  g_{t\phi_i} \simeq 2\Big(-\frac{n}{2k-1} +\frac{2mn}{r^{2k-1}}\Big)
 \cos\theta_i +\cdots \, , \quad
\mathbb{A}_{\phi_i} \simeq \sqrt{k}\frac{2qn}{r^{2k-1}}\cos\theta_i{} +\cdots \, . \nn
\eea

In addition to these primary hairs, one can note that besides the previously introduced
secondary hair $J_n \propto{} mn$ that appears in the next-to-leading order of the asymptotic
expansions of the metric components $g_{t\phi_i}$, there is also a similar quantity $qn$
displayed in the next-to-leading order of the asymptotic expansions of the Abelian potential's
components $\mathbb{A}_{\phi_i}$, indicating that the secondary hair $Q_n \propto{} qn$
might also play a crucial role in the mass formulas.

In order to derive the first law, we now adopt the method used in Refs. \cite{2209.01757,
PLB608-251} to deduce a meaningful squared-mass formula. To do so, we may rewrite the
horizon equation $f(r_h) = 0$ in which $f(r)$ is given by Eq. (\ref{fg0}) with the help
of the reduced horizon area $\cA = \big(r_h^2+n^2\big)^k$ and get the following identity:
\bea\label{sqm}
m^2\cA^{1/k} = (mn)^2 +\frac{1}{4}\Big[X(\cA) +(2k-1)q^2\cA^{1/k -1}
 -2(k-1)\frac{(qn)^2}{\cA}\Big]^2 \, ,
\eea
where
\bea\label{XA}
&& X(\cA) 
= \frac{(\cA^{1/k}-n^2)^{1/2}}{2k}\int^{\cA}\frac{y^{1/k}}{\big(y^{1/k} -n^2\big)^{3/2}}dy \nn \\
&&\qquad\quad = \sum_{i=0}^k \frac{k!\, n^{2i}\big(\cA^{1/k}
 -n^2\big)^{k-i}}{i!(k-i)!\, (2k-2i-1)} \, .
\eea

At this step, a few comments about Eq. (\ref{sqm}) will be in order. First, Eq. (\ref{sqm})
can be viewed a simple deformation of the horizon equation: $f(r_h) = 0$, then it depicts
a hypersurface in the usual four-dimensional state space spanned by the four independent
variables ($\cA, m, n, q$), which exactly match with three solution parameters that appeared
in the function $f(r)$. However, we shall not take this viewpoint as it does not lead to
our expected results, in that the resulted thermodynamic quantities do not constitute the
ordinary conjugate pairs.

On the other hand, Eq. (\ref{sqm}) can be also thought of as representing a hypersurface in
the extended six-dimensional state space spanned by the six variables ($\cA, m, n, q, mn, qn$)
if one regards ($mn, qn$) as two new independent coordinates in the higher-dimensional state
space. These two thermodynamic variables are precisely related to the extra secondary hairs $
J_n \propto{} mn$ and $Q_n \propto{} qn$ that would be introduced in the remaining discussions.
Mathematically, this novel viewpoint is nothing but a kind of the embedding mapping\footnote{
There are many well-known examples of this kind of embedding: A two-sphere can be embedded into
a three-dimensional flat space; A $D$-dimensional pure (anti-)de Sitter spacetime can be embedded
into a ($D +1$)-dimensional flat Minkowski spacetimes. Even the four-dimensional Schwarzschild
black hole can be embedded into a six-dimensional flat spacetime \cite{PR116-778}.}, that is,
we will embed the common four-dimensional state space into a six-dimensional one in which
$J_n$ and $Q_n$ are naturally handled as two independent thermodynamic variables. This unusual
perspective somewhat exhibits a taste of holography in the thermodynamic state space. According
to this new viewpoint, thermodynamics of the NUT-charged (AdS) spacetime can be viewed as being
treated within two higher dimensional state space. In one word, since the inclusion of two extra
secondary hairs $J_n$ and $Q_n$ is a kind of the embedding mapping, the mismatch of the numbers
of parameters in two different spaces should not be considered as a mathematical inconstancy.

Next, one may argue that it is possible to introduce so many different hairs that the resulting
thermodynamical expressions are not unique. Indeed mathematically, there are many possible
manners for embedding a low-dimensional curved space into a higher-dimension flat space, but
there is only one least embedding manner. Although one can introduce many different extra hairs
to do the same job, the least cost to pay is that one should choose the least and most physically
reasonable hair as possible as one could.

Now we are ready to derive the consistent thermodynamic from the above squared-mass formula
(\ref{sqm}) which is regarded formally as a basic functional relation: $m = m(\cA, q, n, mn, qn)$.
Similar to the procedure as did before in Refs. \cite{2209.01757,PRD100-101501,PRD105-124013,
2210.17504}, we can differentiate this equation with respect to its five variables ($\cA, q, n,
mn, qn$) to yield their corresponding conjugate quantities, and subsequently we arrive at the
differential and integral mass formulas with the conjugate thermodynamic potentials being given
by the ordinary Maxwell relations. Obviously, the differentiation of the squared-mass formula
(\ref{sqm}) leads to the first law:
\bea
&& dm = \frac{\p{}m}{\p\cA} d\cA +\frac{\p{}m}{\p{}(mn)} d(mn) +\frac{\p{}m}{\p{}(qn)} d(qn)
 +\frac{\p{}m}{\p{}q} dq +\frac{\p{}m}{\p{}n} dn \nn \\
&&\quad~ = \kappa{}/(2k)d\cA +\omega_h{}d(mn) +\Psi_{qn}d(qn) +\Phi_q{}dq +\psi_h{}dn \, ,
\label{dmg0}
\eea
where the expression of $\kappa = 2k\p{}m/\p\cA$ completely coincides with that given in
Eq. (\ref{kappag0}), and
\bea\label{Therm}
&& \omega_h = \frac{n}{r_h^2+n^2} \, , \quad
 \Phi_q = \frac{(2k-1)qr_h}{\big(r_h^2+n^2\big)^k} \, , \quad
 \Psi_{qn} = \frac{-2(k-1)qnr_h}{\big(r_h^2+n^2\big)^{k+1}} \, , \nn \\
&& \psi_h = -\frac{\big(r_h^2+n^2\big)^k}{2nr_h} +\frac{(2k-1)r_h^2
 -n^2}{2n\big(r_h^2+n^2\big)}\int^{r_h}\frac{ \big(x^2+n^2\big)^k}{x^2}dx \, .
\eea
One can check that they also simultaneously fulfil the integral mass formulas:
\be
(2k-1)(m -q\Phi_q) = \kappa\cA +2k\omega_h{}(mn) +2k\Psi_{qn}(qn) +\psi_h{}n \, , \label{Img0}
\ee
which implies that these quantities indeed constitute the ordinary thermodynamic pairs. By the
way, we also point out that the following identity:
\be
m = \int^{r_h}\frac{\big(x^2+n^2\big)^k}{2x^2}dx
  +q^2\frac{(2k-1)r_h^2 +n^2}{2r_h\big(r_h^2+n^2\big)^k} \, ,
\ee
must be used to verify that both mass formulas are satisfied.

Now one needs further to make the following identifications with the thermodynamical quantities
given before
\bea\label{therm}
&& M = k(4\pi)^{k-1}m \, , \quad  N = k(4\pi)^{k-1}n \, , \quad
Q = \sqrt{k}(2k-1)(4\pi)^{k-1}q \, , \nn \\
&& J_n = k(4\pi)^{k-1}(mn) \, , \quad  Q_n = k(4\pi)^{k-1}(qn) \, , \quad
\Phi = \frac{\sqrt{k}}{2k-1}\Phi_q \, ,
\eea
then Eqs. (\ref{dmg0}) and (\ref{Img0}) are nothing more than the conventional forms of
the first law and the Bekenstein-Smarr mass formula:
\bea
&&\qquad\qquad dM = TdS +\omega_h{}dJ_n +\psi_h{}dN +\Phi{}dQ +\Psi_{qn}dQ_n \, , \\
&& (D-3)(M -Q\Phi) = (D-2)(TS +\omega_h{}J_n +\Psi_{qn}Q_n) +\psi_h{}N \, .
\eea
They obviously show that the higher even-dimensional RN-NUT spacetimes obey the familiar
(extended) forms of differential and integral mass formulas of the usual black hole
thermodynamics, just like the case of the four-dimensional RN-NUT spacetimes as shown
in our previous papers \cite{PRD100-101501,PRD105-124013}. By the way, it is interesting
to note that $\Psi_{qn} = 0$ when $k = 1$, thus demonstrating that one needn't to introduce
the secondary hair $Q_n = Qn = qn$ in the four-dimensional NUT-charged spacetime at all.

The above differential and integral mass formulas strongly suggest that the following
familiar identifications be made:
\be
S = \frac{1}{4}A_h = \frac{1}{4}(4\pi)^k\cA \, , \qquad
T = \frac{\kappa}{2\pi} = \frac{f^{\prime}(r_h)}{4\pi} \, ,
\ee
which naturally recovers the famous Bekenstein-Hawking one-quarter area-entropy relation
for all the ($2k+2$)-dimensional RN-NUT spacetimes.

\section{Mass formulas of ($2k+2$)-dimensional RN-NUT AdS spacetimes}

In this section, we shall extend the preceding discussion to the general ($2k+2$)-dimensional
RN-NUT AdS cases, in which the metric and the Abelian gauge potential are still given by Eqs.
(\ref{hdNUT})-(\ref{mathbbA}), where the radial function can now be written in another more
transparent way:
\bea
&& f(r)= g_0^2\big(r^2+n^2\big) -\frac{2mr}{\big(r^2+n^2\big)^k}
 +q^2\frac{(2k-1)r^2 +n^2}{\big(r^2+n^2\big)^{2k}} \nn \\
&&\qquad\quad +\frac{\big[1 +2(k+1)g_0^2n^2\big]r}{\big(r^2+n^2\big)^k}\int^r
\frac{\big(x^2+n^2\big)^k}{x^2}dx \, . \label{fg}
\eea

Just like the case that was treated in the last section without a cosmological constant, using
the reduced horizon area $\cA = \big(r_h^2+n^2\big)^k$, the horizon equation $f(r_h) = 0$ with
$f(r)$ being given by Eq. (\ref{fg}) can be rewritten as
\bea\label{SQM}
&& m^2\cA^{1/k} = (mn)^2 +\frac{1}{4}\Big\{g_0^2\cA^{1/k +1}
 +\big[1 +2(k+1)g_0^2n^2\big]X(\cA) \nn \\
&&\qquad\qquad\quad +(2k-1)q^2\cA^{1/k -1} -2(k-1)\frac{(qn)^2}{\cA}\Big\}^2 \, ,
\eea
where the expression of $X(\cA)$ is still given by Eq. (\ref{XA}).

Eq. (\ref{SQM}) represents a hypersurface in the five-dimensional state space spanned by the
coordinates ($\cA, m, n, q, g_0$). The reduced forms of the first law and Bekenstein-Smarr mass
formula that are derived directly from this perspective don't consist of the usual conjugate
pairs about their corresponding thermodynamic quantities. Therefore, similar to the strategy as
did in the last section, we shall also import two extra secondary hairs $J_n \propto{} mn$
and $Q_n \propto{} qn$ so that Eq. (\ref{SQM}) now depicts a hypersurface in the extended
seven-dimensional state space spanned by the thermodynamic variables ($\cA, m, n, q, mn, qn,
g_0$), from which one can view the mass parameter $m$ as an implicit function: $m = m(\cA, n,
q, mn, qn, g_0)$. In this way, $mn$ and $qn$ should now be regarded as two new independent
coordinates in the seven-dimensional thermodynamical state space.

Take this opinion and parallel to the steps in the last section, one can differentiate the
squared-mass formula (\ref{SQM}) with respect to its seven variables ($\cA, m, n, q, mn, qn,
g_0$) to obtain the following differential and integral mass formulas:
\bea
&& dm = \kappa{}/(2k)d\cA +\omega_h{}d(mn) +\Psi_{qn}d(qn) +\Phi_q{}dq
 +\psi_h{}dn +\Theta{}d(g_0^2) \, , \qquad \\
&& (2k-1)(m -q\Phi_q) = \kappa\cA +2k\omega_h{}(mn) +2k\Psi_{qn}(qn)
 +\psi_h{}n -2g_0^2\Theta \, ,
\eea
where the expression of $\kappa = 2k\p{}m/{\p\cA}$ reproduces the result given in Eq. (\ref{kath}),
while those of $\omega_h, \Phi_q, \Psi_{qn}$ are the same ones as those given by the first line
of Eq. (\ref{Therm}), and two new expressions are
\bea
&& \psi_h = -\frac{1+2(k+1)g_0^2n^2}{2nr_h}\big(r_h^2+n^2\big)^k
 +\Big[\frac{k +2(k+1)^2g_0^2n^2}{n\big(r_h^2+n^2\big)}r_h^2 \nn \\
&&\qquad -\frac{1+2(k+1)g_0^2n^2}{2n}\Big]\int^{r_h}\frac{(x^2+n^2)^k}{x^2}dx \, , \nn \\
&& \Theta = \frac{\p{}m}{\p{}(g_0^2)} = \frac{(2k+1)r_h^2}{2\big(r_h^2+n^2\big)}
 \int^{r_h}\frac{\big(x^2+n^2\big)^{k+1}}{x^2}dx \nn \\
&&\quad = \frac{r_h}{2}\big(r_h^2+n^2\big)^k
 +\frac{(k+1)n^2r_h^2}{r_h^2+n^2}\int^{r_h}\frac{\big(x^2+n^2\big)^k}{x^2}dx \, .
\eea
By the way, the following identity must be used to verify that both mass formulas are indeed
fulfilled:
\bea
&& m = \int^{r_h}\big[1 +(2k+1)g_0^2\big(x^2+n^2\big)\big]
 \frac{\big(x^2+n^2\big)^k}{2x^2}dx
  +q^2\frac{(2k-1)r_h^2 +n^2}{2r_h\big(r_h^2+n^2\big)^k} \nn \\
&&\quad = \frac{1 +2(k+1)g_0^2n^2}{2}\int^{r_h}\frac{\big(x^2+n^2\big)^k}{x^2}dx
 +\frac{g_0^2}{2r_h}\big(r_h^2+n^2\big)^{k+1} \nn \\
&&\qquad +q^2\frac{(2k-1)r_h^2 +n^2}{2r_h\big(r_h^2+n^2\big)^k} \, .
\eea

Therefore, we have revealed that the higher even-dimensional RN-NUT-AdS spacetimes can also
comply with the traditional (but extended) forms of the first law and the Bekenstein-Smarr
mass formula as follows:
\bea
dM = TdS +\omega_h{}dJ_n +\psi_h{}dN +\Phi{}dQ +\Psi_{qn}dQ_n +VdP \, , && \\
(D-3)(M -Q\Phi) = (D-2)(TS +\omega_h{}J_n +\Psi_{qn}Q_n) +\psi_h{}N -2VP \, , &&
\eea
provided that two new secondary hairs: $J_n = Mn$ and $Q_n \propto qn$, and also the pressure
\cite{PRD84-024037} and thermodynamic volume are simultaneously introduced:
\be
P = \frac{k(2k+1)}{8\pi}g_0^2 \, , \qquad  V = 2\frac{(4\pi)^k}{2k+1}\Theta \, .
\ee
In addition, it should be noted that the thermodynamical quantities that enter the above
differential and integral mass formulas are still given by Eq. (\ref{therm}) for ($M, N, Q,
J_n, Q_n, \Phi$). However, $\Psi_{qn}$ vanishes when $k = 1$, indicating that it need not
to be introduced in the case of the four-dimensional RN-NUT AdS cousin \cite{PRD100-101501}.
Finally, when the charge parameter vanishes ($q = 0$), all the above results reproduce those
presented in our previous work \cite{2209.01757}.

Subsequently, it is highly urged that we must have the conclusions: $S = A_h/4\, , T = \kappa/
(2\pi)$, which naturally recovers the famous Bekenstein-Hawking one-quarter area-entropy law
and Hawking-Gibbons relation of all the ($2k+2$)-dimensional RN-NUT-AdS spacetimes.

\section{Conclusions and outlooks}

In this paper, we have generalized our previous work \cite{2209.01757} to derive the consistent
thermodynamic first law and Bekenstein-Smarr mass formula for all higher even-dimensional RN-NUT
(AdS) spacetimes provided that we introduce a new secondary hair $Q_n \propto{} qn$ in addition to
$J_n = Mn$. By the way, we have also shown that there is no need to introduce the secondary hair
$Q_n = Qn = qn$ in the mass formulas for the four-dimensional purely electric-charged RN-NUT (AdS)
spacetimes. What is more, we have also demonstrated that all formalisms that are based upon the
(modified) Komar integral are unlikely to be directly applied in all $D > 4$ dimensional (RN-)
NUT-charged (AdS) spacetimes, their success in the four-dimensional NUT-charged (AdS) spacetimes
is only an accident and coincidence event.

The study in this paper further approves that our ``multi-hair" formalism is a universal and
successful framework to deal with the consistent thermodynamics of all the NUT-charged spacetimes.
In this scheme, the squared-mass formula is interpreted as representing a hypersurface in the
higher dimensional thermodynamical state space after some extra secondary hairs are introduced,
which is nothing more than a kind of the embedding mapping that reflects the spirt of holography
in the thermodynamical aspects. In the higher-dimensional thermodynamical state space, our first
law and the Bekenstein-Smarr mass formula acquire their familiar forms of the usual black hole
thermodynamics, and the related thermodynamic quantities not only obey the common Maxwell
relations but also constitute the usual conjugate pairs. Without the inclusion of these new
extra secondary hairs, these mass formulas will boil down to the forms in which all perfect
relations are lost.

There are two possible directions along the present work. A related subject is how to extend our
formalism to copy with the consistent thermodynamics of the higher even-dimensional multi-NUTty
spacetimes \cite{CQG21-2937,PLB634-448}. Another potential application of our scheme is to
investigate the thermodynamics of the Lorentzian NUT-charged solutions to other modified
gravity theories, such as Gauss-Bonnet, Lovelock, $f(R)$ and $f(T)$ theories, etc.

\section*{Acknowledgments}
We are greatly indebted to the anonymous referee for his/her invaluable comments to improve the
presentations of this work. This work is supported by the National Natural Science Foundation of
China (NSFC) under Grant No. 12205243, No. 12375053, No. 11675130, by the Sichuan Science and
Technology Program under Grant No. 2023NSFSC1347, and by the Doctoral Research Initiation Project
of China West Normal University under Grant No. 21E028.

\appendix
\setcounter{equation}{0}

\section{(Alternative) method to compute the
mass and secondary hairs: $J_n$ and $Q_n$}\label{appe}

The main purpose of this appendix is to suggest a universal definition to evaluate the secondary
hairs: $J_n$ and $Q_n$. In addition, we will also supplement a massless background subtracted Komar
integral to get the finite mass expressions. All these approaches work effectively in all (even)
dimensions.

In the main context, we have adopted the AD method to compute the mass in all even-dimensional
NUT-charged spacetimes, both for the cases without a cosmological constant and with a nonzero
one. We also point out that in the cases of the AdS-NUT spacetimes, the counter-term mass is
identical to the AD mass. These are the only effective approaches to compute the conserved mass
of all high even-dimensional NUT-charged spacetimes. As a supplement, here we propose to utilize
the massless background subtracted Komar integral to evaluate the mass for all even-dimensional
NUT-charged (AdS) spacetimes:
\bea
M = -\frac{D-2}{16(D-3)\pi}\int ({^\star}\Xi -{^\star}\bar{\Xi})
 = \frac{-k}{8(2k-1)\pi}\int_{\bigotimes^k{}S_{\infty}^2}
 \sqrt{-g}(\Xi^{tr} -\bar{\Xi}^{tr})d^{2k}x \, , \label{bsKm}
\eea
where a star represents the Hodge dual, and the 2-form $\Xi = d\xi$ is the usual Komar super-potential:
\be
\Xi = -\p_r{}f(r){} dr\wedge \Big(dt +2n\sum_{i=1}^k\cos\theta_i{}d\phi_i\Big)
 +2nf(r)\sum_{i=1}^k\sin\theta_i{}d\theta_i \wedge{} d\phi_i \, ,
\ee
associated with the timelike vector $\xi^a = (\p_t)^a$, whose dual 1-form is: $\xi = -f(r)\big(dt
+2n\sum_{i=1}^k\cos\theta_i{}d\phi_i\big)$. The related background 1-form is: $\bar{\xi} = -\bar{f}(r)
\big(dt +2n\sum_{i=1}^k\cos\theta_i{}d\phi_i\big)$, in which $\bar{f}(r) = f(r)|_{m=0}$.

Denoting $m(r) = (r^2+n^2)^k\p_r{}f(r)$ in the mass function $\mathcal{M}(r)$ given by Eq. (\ref{Mr}),
one can observe that the large-$r$ asymptotic behaviors of $m(r)$ and $f(r)$ in the $D = 4, 6, 8, 10$
are as follows:
\bea
m_4 &=& 2g_0^2r\big(r^2+n^2\big) +2m +\mathcal{O}(r^{-1}) \, , \nn \\
m_6 &=& 2g_0^2r\big(r^2+n^2\big)^2 -\frac{8}{3}\big(1 +6g_0^2n^2\big)n^2r +6m +\mathcal{O}(r^{-1}) \, ,
\nn \\
m_8 &=& 2g_0^2r\big(r^2+n^2\big)^3 -\frac{4}{5}\big(1 +8g_0^2n^2\big)n^2r\big(r^2 +9n^2\big)
 +10m +\mathcal{O}(r^{-1}) \, , \nn \\
m_{10} &=& 2g_0^2r\big(r^2+n^2\big)^4 -\frac{16}{35}\big(1 +10g_0^2n^2\big)n^2r
 \big(r^4 +6n^2r^2 +29n^4\big) +14m +\mathcal{O}(r^{-1}) \, , \nn
\eea
and
\bea
f_4 &=& g_0^2r^2 +1 +5g_0^2n^2 -\frac{2m}{r} +\mathcal{O}(r^{-2}) \, , \nn \\
f_6 &=& g_0^2r^2 +\frac{1 +9g_0^2n^2}{3} +\frac{4}{3}\big(1 +6g_0^2n^2\big)n^2\frac{1}{r^2}
 -\frac{2m}{r^3} +\mathcal{O}(r^{-4}) \, , \nn \\
f_8 &=& g_0^2r^2 +\frac{1 +13g_0^2n^2}{5} +\frac{2}{5}\big(1 +8g_0^2n^2\big)n^2\frac{r^2+3n^2}{r^4}
 -\frac{2m}{r^5} +\mathcal{O}(r^{-6}) \, , \nn \\
f_{10} &=& g_0^2r^2 +\frac{1 +17g_0^2n^2}{7} +\frac{8}{35}\big(1 +10g_0^2n^2)n^2
\frac{r^4 +n^2r^2 +5n^4}{r^6} -\frac{2m}{r^7} +\mathcal{O}(r^{-8}) \, . \nn
\eea
So using the massless purely NUT-charged spacetimes as the reference backgrounds and with the help
of determinant: $\sqrt{-g} = (r^2+n^2)^k\prod_{i=1}^k\sin\theta_i$, the above background subtracted
Komar integral (\ref{bsKm}) precisely gives the expected expressions for the mass: $M = k(4\pi)^{k-1}m$.
Note also that we have included a dimension-dependent factor $(D-2)/(D-3) = 2k/(2k-1)$ to recover
the correct coefficient. In the meanwhile, the above result also indicates that one can use this
background subtracted Komar integral to derive the Bekenstein-Smarr mass formula.

Now we would like to tentatively define the NUT charge $N$ and the secondary hairs $J_n$ and $Q_n$
in all diverse dimensions. Since the results differ from those ones in the main context by a common
factor $k(4\pi)^{k-1}$, so we will denote them by a hat to make a distinction. Likewise in the
four-dimensional case \cite{PRD59-024009}, the $i$-th NUT charge is defined as:
\be
\hat{N}^i = \frac{-1}{8\pi}\int_{S_i^2}d\hat{\xi}
 = \frac{n}{2}\int_0^{\pi} \sin\theta_i d\theta_i = n \, ,
\ee
where $\hat{\xi} = \xi/g_{tt} = dt +2n\sum_{i=1}^k\cos\theta_i{}d\phi_i$.

The $i$-th secondary hair $J_n^i$ and $Q_n^i$ are, respectively, defined as:
\bea
\hat{J}_n^i &=& \frac{-1}{16\pi}\int_{S_i^2}r^{D-3}(\Xi -\bar{\Xi})
 = \frac{mn}{2}\int_0^{\pi} \sin\theta_i d\theta_i = mn \, , \\
\hat{Q}_n^i &=& \frac{-1}{8\pi}\int_{S_i^2}r^{D-3}F
 = \frac{qn}{2}\int_0^{\pi} \sin\theta_i d\theta_i = qn \, ,
\eea
where the Maxwell 2-form $F = d\mathbb{A}$ reads
\be
F = \sqrt{k}q\frac{(1-2k)r^2+n^2}{\big(r^2 +n^2\big)^{k+1}} dr\wedge \Big(dt
 +2n\sum_{i=1}^k\cos\theta_i{}d\phi_i\Big) -\frac{2\sqrt{k}qnr}{\big(r^2 +n^2\big)^k}
\sum_{i=1}^k\sin\theta_i{} d\theta_i \wedge{} d\phi_i \, .
\ee
Apart from the need of the massless background substraction, the factor $r^{D-3} = r^{2k-1}$
clearly demonstrates that they are the sub-leading dual charges.

The above definitions work also effectively in the multi-NUTty cases. However, since there is no
exact multi-NUTty solution so far being found in the Einstein-Maxwell theory, the definition for
$Q_n^i$ is only formal, but may be extended to other super-gravity theories.

It is an open issue that whether the secondary hairs introduced in our papers can be understood
as some subleading dual charges in terms of the asymptotic Bondi-Metzner-Sachs (BMS) algebras or
the Newman-Penrose charges \cite{JHEP0119143,JHEP0319057}. The above definitions need a further
justification along this line.


\begin{thebibliography}{99}

\def\CQG{Classical Quantum Gravity \,}
\def\JHEP{J. High Energy Phys. \,}
\def\PRD{Phys. Rev. D \,}
\def\PRL{Phys. Rev. Lett. \,}
\def\NPB{Nucl. Phys. B}
\def\PLB{Phys. Lett. B \,}

\bibitem{PRD100-101501}
S.-Q. Wu and D. Wu,
\textrm{Thermodynamical hairs of the four-dimensional Taub-Newman-Unti-Tamburino spacetimes},
\href{https://doi.org/10.1103/PhysRevD.100.101501}
{\PRD \textbf{100}, 101501(R) (2019)}.

\bibitem{PRD105-124013}
D. Wu and S.-Q. Wu,
\textrm{Consistent mass formulas for the four-dimensional dyonic NUT-charged spacetimes},
\href{https://doi.org/10.1103/PhysRevD.105.124013}
{\PRD \textbf{105}, 124013 (2022)}.

\bibitem{2210.17504}
D. Wu and S.-Q. Wu,
\textrm{Revisiting mass formulas of the four-dimensional Reissner-Nordstrom-NUT-AdS solutions
in a different metric form},
\href{https://arxiv.org/abs/2210.17504}{arXiv:2210.17504 [gr-qc]}.

\bibitem{PRD100-064055}
R.A. Hennigar, D. Kubiz\v{n}\'ak, and R.B. Mann,
\textrm{Thermodynamics of Lorentzian Taub-NUT spacetimes},
\href{https://doi.org/10.1103/PhysRevD.100.064055}
{\PRD \textbf{100}, 064055 (2019)}.

\bibitem{JHEP0719119}
A.B. Ballon, F. Gray, and D. Kubiz\v{n}\'ak,
\textrm{Thermodynamics and phase transitions of NUTty dyons},
\href{https://doi.org/10.1007/JHEP07(2019)119}
{\JHEP \textbf{07} (2019) 119}.

\bibitem{CQG36-194001}
A.B. Bordo, F. Gray, R.A. Hennigar, and D. Kubiz\v{n}\'ak,
\textrm{Misner gravitational charges and variable string strengths},
\href{https://doi.org/10.1088/1361-6382/ab3d4d}
{\CQG \textbf{36}, 194001 (2019)}.

\bibitem{PLB798-134972}
A.B. Bordo, F. Gray, R.A. Hennigar, and D. Kubiz\v{n}\'ak,
\textrm{The first law for rotating NUTs},
\href{https://doi.org/10.1016/j.physletb.2019.134972}
{\PLB \textbf{798}, 134972 (2019)}.

\bibitem{PRD100-104016}
Z.H. Chen and J. Jiang,
\textrm{General Smarr relation and first law of a NUT dyonic black hole},
\href{https://doi.org/10.1103/PhysRevD.100.104016}
{\PRD \textbf{100}, 104016 (2019)}.

\bibitem{JHEP0520084}
A.B. Ballon, F. Gray, and D. Kubiz\v{n}\'ak,
\textrm{Thermodynamics of rotating NUTty dyons},
\href{https://doi.org/10.1007/JHEP05(2020)084}
{\JHEP \textbf{05} (2020) 084}.

\bibitem{JHEP1022044}
N.H. Rodriguez and M.J. Rodriguez,
\textrm{First law for Kerr Taub-NUT AdS black holes},
\href{https://doi.org/10.1007/JHEP10(2022)044}
{\JHEP \textbf{10} (2022) 044}.

\bibitem{IJMPD31-2250021}
R. Durka,
\textrm{The first law of black hole thermodynamics for Taub-NUT spacetime},
\href{https://doi.org/10.1142/S0218271822500213}
{Int. J. Mod. Phys. D \textbf{31}, 2250021 (2022)}.

\bibitem{PLB802-135270}
G. Cl\'ement and D. Gal'tsov,
\textrm{On the Smarr formulas for electrovac spacetimes with line singularities},
\href{https://doi.org/10.1016/j.physletb.2020.135270}
{\PLB \textbf{802}, 135270 (2020)}.

\bibitem{PRD106-024022}
M. Godazgar and S. Guisset,
\textrm{Dual charges for AdS spacetimes and the first law of black hole mechanics},
\href{https://doi.org/10.1103/PhysRevD.106.024022}
{\PRD \textbf{106}, 024022 (2022)}.

\bibitem{JHEP0321039}
R.B. Mann, L.A.P. Zayas, and M. Park,
\textrm{Complement to thermodynamics of dyonic Taub-NUT-AdS spacetime},
\href{https://doi.org/10.1007/JHEP03(2021)039}
{\JHEP \textbf{03} (2021) 039}.

\bibitem{PRD101-124011}
A.M. Awad and S. Eissa,
\textrm{Topological dyonic Taub-Bolt/NUT-AdS solutions: Thermodynamics and first law},
\href{https://doi.org/10.1103/PhysRevD.101.124011}
{\PRD \textbf{101}, 124011 (2020)}.

\bibitem{PRD105-124034}
A.M. Awad and S. Eissa,
\textrm{Lorentzian Tuab-NUT spacetimes: Misner string charges and the first law},
\href{https://doi.org/10.1103/PhysRevD.105.124034}
{\PRD \textbf{105}, 124034 (2022)}.

\bibitem{JHEP1022174}
H.-S. Liu, H. L\"{u}, and L. Ma,
\textrm{Thermodynamics of Taub-NUT and Plebanski solutions},
\href{https://doi.org/10.1007/JHEP10(2022)174}
{\JHEP \textbf{10} (2022) 174}.

\bibitem{2209.01757}
D. Wu and S.-Q. Wu,
\textrm{Consistent mass formulas for higher even-dimensional Taub-NUT spacetimes
and their AdS counterparts},
\href{https://doi.org/10.1103/PhysRevD.108.064034}
{\PRD \textbf{108}, 064034 (2023)}.

\bibitem{NPB195-76}
L.F. Abbott and S. Deser,
\textrm{Stability of gravity with a cosmological constant},
\href{https://doi.org/10.1016/0550-3213(82)90049-9}
{\NPB \textbf{195}, 76 (1982)}.

\bibitem{PRD73-064020}
H. Cebeci, O. Sarioglu, and B. Tekin,
\textrm{Negative mass solitons in gravity},
\href{https://doi.org/10.1103/PhysRevD.73.064020}
{\PRD \textbf{73}, 064020 (2006)}.

\bibitem{CQG23-2849}
A.M. Awad,
\textrm{Higher dimensional Taub-NUTs and Taub-bolts in Einstein-Maxwell gravity},
\href{https://doi.org/10.1088/0264-9381/23/9/006}
{\CQG \textbf{23}, 2849 (2006)}.

\bibitem{PLB632-537}
R.B. Mann and C. Stelea,
\textrm{New Taub-NUT-Reissner-Nordstrom spaces in higher dimensions},
\href{https://doi.org/10.1016/j.physletb.2005.10.085}
{\PLB \textbf{632}, 537 (2006)}.

\bibitem{PRD73-124039}
M.H. Dehghani and A. Khodam-Mohammadi,
\textrm{Thermodynamics of Taub-NUT black holes in Einstein-Maxwell gravity},
\href{https://doi.org/10.1103/PhysRevD.73.124039}
{\PRD \textbf{73}, 124039 (2006)}.

\bibitem{hepth9809041}
M. Taylor,
\textrm{Higher dimensional Taub-Bolt solutions and the entropy of noncompact
manifolds},
\href{https://arxiv.org/abs/hep-th/9809041}{hep-th/9809041}.

\bibitem{CQG19-2051}
A.M. Awad and A. Chamblin,
\textrm{A bestiary of higher dimensional Taub-NUT AdS space-times},
\href{https://doi.org/10.1088/0264-9381/19/8/301}
{\CQG \textbf{19}, 2051 (2003)}.

\bibitem{PLB593-218}
H. L\"{u}, D.N. Page, and C.N. Pope,
\textrm{New inhomogeneous Einstein metrics on sphere bundles over Einstein-Kahler manifolds},
\href{https://doi.org/10.1016/j.physletb.2004.04.068}
{\PLB \textbf{593}, 218 (2004)}.

\bibitem{PRD60-104001}
R. Emparan, C.V. Johnson, and R.C. Myers,
\textrm{Surface terms as counterterms in the AdS/CFT correspondence},
\href{https://doi.org/10.1103/PhysRevD.60.104001}
{\PRD \textbf{60}, 104001 (1999)}.

\bibitem{PRD72-083032}
R.B. Mann and C. Stelea,
\textrm{On the thermodynamics of NUT charged spaces},
\href{https://doi.org/10.1103/PhysRevD.72.084032}
{\PRD \textbf{72}, 084032 (2005)}.

\bibitem{PLB620-1}
D. Astefanesei, R. B. Mann, and E. Radu,
\textrm{Breakdown of the entropy/area relationship for NUT-charged spacetimes},
\href{https://doi.org/10.1016/j.physletb.2005.05.057}
{\PLB \textbf{620}, 1 (2005)}.

\bibitem{NPB652-348}
R. Clarkson, L. Fatibene, and R.B. Mann,
\textrm{Thermodynamics of (d+1)-dimensional NUT-charged AdS spacetimes},
\href{https://doi.org/10.1016/S0550-3213(02)01143-4}
{\NPB \textbf{652}, 348 (2003)}.

\bibitem{NPB674-329}
R. Clarkson, A.M. Ghezelbash, and R.B. Mann,
\textrm{Entropic N-bound and maximal mass conjecture violations in four
dimensional Taub-bolt(NUT)-dS spacetimes},
\href{https://doi.org/10.1016/j.nuclphysb.2003.09.039}
{\NPB \textbf{674}, 329 (2003)}.

\bibitem{PRL91-061031}
R. Clarkson, A.M. Ghezelbash, and R.B. Mann,
\textrm{Mass, Action, and Entropy of Taub-Bolt-de Sitter Spacetimes},
\href{https://doi.org/10.1103/PhysRevLett.91.061301}
{\PRL \textbf{91}, 061301 (2003)}.

\bibitem{IJMPA19-3987}
R. Clarkson, A.M. Ghezelbash, and R.B. Mann,
\textrm{A review of the N-bound and the maximal mass conjectures using NUT-charged
dS spacetimes},
\href{https://doi.org/10.1142/S0217751X04019822}
{Int. J. Mod. Phys. A \textbf{19}, 3987 (2004)}.

\bibitem{JHEP0105049}
D. Astefanesei, R.B. Mann, and E. Radu,
\textrm{Nut charged space-times and closed timelike curves on the boundary},
\href{https://doi.org/10.1088/1126-6708/2005/01/049}
{\JHEP \textbf{01} (2005) 049}.

\bibitem{JHEP0306083}
R.-G. Cai and L.-M. Cao,
\textrm{Conserved charges in even dimensional asymptotically locally anti-de
Sitter space-times},
\href{https://doi.org/10.1088/1126-6708/2006/03/083}
{\JHEP \textbf{03} (2006) 083}.

\bibitem{PLB738-294}
C.Oh Lee,
\textrm{The extended thermodynamic properties of Taub-NUT/bolt-AdS spaces},
\href{https://doi.org/10.1016/j.physletb.2014.09.046}
{\PLB \textbf{738}, 294 (2014)}.

\bibitem{JHEP0804045}
D. Kastor and J. Traschen,
\textrm{Conserved gravitational charges from Yano tensors},
\href{https://doi.org/10.1088/1126-6708/2004/08/045}
{\JHEP \textbf{08} (2004) 045}.

\bibitem{PLB608-251}
S.-Q. Wu,
\textrm{New formulation of the first law of black hole thermodynamics: A stringy analogy},
\href{https://doi.org/10.1016/j.physletb.2005.01.018}
{\PLB \textbf{608}, 251 (2005)}.

\bibitem{PR116-778}
C. Fronsdal,
\textrm{Completion and embedding of the Schwarzschild solution},
\href{https://doi.org/10.1103/PhysRev.116.778}
{Phys. Rev. \textbf{116}, 778 (1959)}.

\bibitem{PRD84-024037}
M. Cveti\v{c}, G.W. Gibbons, D. Kubiz\v{n}\'ak, and C.N. Pope,
\textrm{Black hole enthalpy and an entropy inequality for the thermodynamic volume},
\href{https://doi.org/10.1103/PhysRevD.84.024037}
{\PRD \textbf{84}, 024037 (2011)}.

\bibitem{CQG21-2937}
R.B. Mann and C. Stelea,
\textrm{Nuttier (A)dS black holes in higher dimensions},
\href{https://doi.org/10.1088/0264-9381/21/12/010}
{\CQG \textbf{21}, 2937 (2004)}.

\bibitem{PLB634-448}
R.B. Mann and C. Stelea,
\textrm{New multiply nutty spacetimes},
\href{https://doi.org/10.1016/j.physletb.2006.02.019}
{\PLB \textbf{634}, 448 (2006)}.

\bibitem{PRD59-024009}
C.J. Hunter,
\textrm{Action of instantons with nut charge},
\href{https://doi.org/10.1103/PhysRevD.59.024009}
{\PRD \textbf{59}, 024009 (1999)}.

\bibitem{JHEP0119143}
H. Godazgar, M. Godazgar, and C. N. Pope,
\textrm{Subleading BMS charges and fake news near null infinity},
\href{https://doi.org/10.1007/JHEP01(2019)143}
{\JHEP \textbf{01} (2019) 143}.

\bibitem{JHEP0319057}
H. Godazgar, M. Godazgar, and C.N. Pope,
\textrm{Tower of subleading dual BMS charges},
\href{https://doi.org/10.1007/JHEP03(2019)057}
{\JHEP \textbf{03} (2019) 057}.

\end{thebibliography}
\end{document}